\documentclass[12pt]{article}

\usepackage{amssymb}
\usepackage{amsmath}
\usepackage{latexsym}

\catcode`@=11
\renewcommand{\section}{\@startsection{section}{1}{0pt}{\medskipamount}
{\medskipamount}{\large\bf}}
\numberwithin{equation}{section}
\catcode`@=12

\def\beq{\begin{eqnarray}}    
\def\eeq{\end{eqnarray}}      

\begin{document}

\begin{center}
{\Large\bf On multiplicative renormalizability of Yang-Mills theory
with the background field method in the BV-formalism}

\vspace{18mm}

{\Large Igor A. Batalin$^{(a,c)}\footnote{E-mail: batalin@lpi.ru}$\;,
 Klaus Bering$^{(b)}\footnote{E-mail: bering@physics.muni.cz}$\;,
Peter M. Lavrov$^{(c,d)}\footnote{E-mail: lavrov@tspu.edu.ru}$\;,
\\ Igor V. Tyutin$^{(a,c)}\footnote{E-mail: tyutin@lpi.ru}$ }

\vspace{8mm}

\noindent ${{}^{(a)}}$
{\em P.N. Lebedev Physical Institute,\\
Leninsky Prospect \ 53, 119 991 Moscow, Russia}

\noindent ${{}^{(b)}}$
{\em Masaryk University, Faculty  of  Science,\\
Kotlarska  2,  611 37  Brno, Czech Republic}

\noindent  ${{}^{(c)}}
${\em
Tomsk State Pedagogical University,\\
Kievskaya St.\ 60, 634061 Tomsk, Russia}

\noindent  ${{}^{(d)}}
${\em
National Research Tomsk State  University,\\
Lenin Av.\ 36, 634050 Tomsk, Russia}

\vspace{20mm}

\begin{abstract}
\noindent Studying the gauge-invariant renormalizability of
four-dimensional  Yang-Mills theory using the background field
method and the BV-formalism, we derive a classical master-equation
homogeneous with respect to the antibracket by introducing antifield
partners to the background fields and parameters.  The constructed
model can be renormalized by the standard method of introducing
counterterms. This model does not have  (exact) multiplicative
renormalizability but it does have   this property  in the physical
sector (quasimultiplicative renormalizability).
\end{abstract}
\end{center}

\vfill


\\

\section{Introduction}

\noindent We analyze the problems of the multiplicative
renormalization  and gauge dependence in the traditional
four-dimensional  Yang-Mills  theory \cite{YM} using the background
field method  \cite{DeW,AFS,Abbott} and the BV formalism
\cite{BV,BV1}. It is well-known that BRST symmetry
\cite{BRS1,T,BRS2}  and the BV formalism \cite{BV,BV1} are
instrumental in proving renormalizability of Yang-Mills theories.

\noindent The basic advantage  of using the background field method
(which introduces a background field $\mathcal{B}_{\mu}$ to the
Yang-Mills gauge field $A_{\mu}$) is that a background gauge
symmetry is preserved after gauge-fixing, thus reducing the list of
possible gauge-invariant counterterms when discussing
renormalization \cite{Barv,BLT-YM1,BLT-YM2}.

\noindent Our main idea here  is to try  to extend antisymplectic
phase space (possibly selectively) by auxiliary sectors. We realize
this idea in the model under consideration, and this procedure can
be generalized to the case with matter. Of course, we cannot yet say
that there is a systematic procedure that works for an arbitrary
gauge theory.

\noindent Although there are many papers devoted to various aspects
of renormalizability of Yang-Mills theories, the gauge dependence of
the renormalization constants has been explicitly studied only in
the gauge field sector \cite{K-SZ}.

\noindent In 1975 Kluberg-Stern and Zuber \cite{K-SZ} introduced a
background field $L_{\mu}(x)$ (which we  call $\theta_{\mu}(x)$) as
a shift part of the BRST transformation of $A_{\mu}$ and a fermionic
$x$-independent parameter $L$ (which we will call $\chi$). The
latter is associated with another kind of shift symmetry for the
BRST transformation. In 1985 Piguet and Sibold \cite{PS} included a
BRST transformation for the $\xi$ parameter (known from the
$R_{\xi}$ gauges). In our previous  papers \cite{BLT-YM1,BLT-YM2} on
the topic (written by three of us)
 in the presence of fermionic and scalar matter,
 the BV-formalism was also not used,
and a generalization of the action from in \cite{K-SZ} was used. But
the corresponding  equations for the classical and effective actions
could not be written in the antibracket form.  This complicates the
cohomological analysis concerning the existence and uniqueness of
solutions.

\noindent Here, we present   an for the action equation in the
homogeneous form of a master equation by introducing antifields to
$\theta_{\mu}$, $\chi$ and $\xi$.

\noindent The paper is organized as follows. In Section
\ref{secbvact} we find a solution  to the master-equation in the
minimal sector and compare it with previous solutions. In Sections
\ref{secssol}-\ref{secpsol} we derive the general solution of the
classical master equation in full antisymplectic phase space.

\section*{Notation}

\noindent We use the condensed DeWitt notation \cite{DeWitt} through
the paper. A left (functional) derivative wrt. a field $\Phi$ is
denoted $\partial_{\Phi}$ while the corresponding right (functional)
derivative is denoted $\overleftarrow{\partial}_{\!\Phi}$.
Space-time indices are written  as subscripts  and denoted by
letters from the middle of the Greek alphabet. Lie algebra indices
are written  as superscripts  and denoted by letters from the
beginning of the Greek alphabet. The Einstein summation convention
is used with a slight modification that an (inverse) metric tensor
is implied for each pair of repeated indices.

\section{BV-action}
\label{secbvact}

\noindent We first consider the traditional four-dimensional
Yang-Mills action  with $SU(2)$ Lie group
\begin{eqnarray}
\mathcal{S}_{YM}(A) &=&\int dx\left(
-\frac{1}{4}G_{\mu \nu }^{\alpha}(A)G_{\mu \nu }^{\alpha }(A)\right),   \label{1.1} \\
G_{\mu \nu }^{\alpha }(A) &=&\partial _{\mu }A_{\nu }^{\alpha }-\partial
_{\nu }A_{\mu }^{\alpha }+g\varepsilon ^{\alpha \beta \gamma }A_{\mu
}^{\beta }A_{\nu }^{\gamma }.
\end{eqnarray}
We here choose $SU(2)$ as the Lie group,   but our construction
works almost unchanged  with any simple compact Lie group. Action
(\ref{1.1}) is invariant $\delta _{\omega }\mathcal{S} _{YM}(A)=0$,
under the transformation
\begin{equation}
\delta _{\omega }A_{\mu }^{\alpha}
=D_{\mu }^{\alpha \beta }(A)\omega^{\beta}, \qquad
D_{\mu }^{\alpha \beta }(A)= \partial_{\mu }\delta^{\alpha\beta }
+ g\varepsilon ^{\alpha \sigma \beta }A_{\mu }^{\sigma } .
\end{equation}
with gauge parameters $\omega^{\alpha }=\omega^{\alpha }(x)$. In the
background field method, we replace $A_{\mu }^{\alpha } \rightarrow
A_{\mu }^{\alpha }+\mathcal{B}_{\mu }^{\alpha }$,
\begin{equation}
\mathcal{S}_{YM}(A)\;\rightarrow \;
\mathcal{S}_{YM}(A+\mathcal{B}),
\end{equation}
where $\mathcal{B}$ is the background field. This action is
invariant under the gauge transformation
\begin{equation}
\delta_{\lambda }A_{\mu }^{\alpha }=\delta A_{\mu }^{\alpha }\lambda ,\qquad
\delta A_{\mu }^{\alpha }=D_{\mu }^{\alpha \beta }(V)C^{\beta },\qquad
V=A+\mathcal{B}.
\end{equation}
Further,  we  follow the BV formalism. The minimal sector consists
the set of fields and antifields
\begin{eqnarray}
&&\Phi =\{A_{\mu }^{\alpha }(x),\ C^{\alpha }(x),\ \mathcal{B}_{\mu
}^{\alpha }(x),\ \theta _{\mu }^{\alpha }(x)\}, \\
&&\Phi^{\ast }=\{A_{\mu }^{\ast \alpha }(x),\ C^{\ast \alpha }(x),\
\mathcal{B}_{\mu }^{\ast \alpha }(x),\ \theta _{\mu }^{\ast \alpha }(x)\}.
\end{eqnarray}
In Table 1,  we present the ''quantum'' numbers of fields,
antifields, auxiliary fields and constant parameters used in
construction of the action. We recall that
\begin{eqnarray}
\textrm{gh}(\Phi^{\ast})=-1-\textrm{gh}(\Phi), \qquad
\textrm{dim}(\Phi^{\ast})=3-\textrm{dim}(\Phi).
\end{eqnarray}

\begin{table}[ht]
\begin{center}
\begin{tabular}{|c||c|c|c||c|c|c|c|c|c||c|c|c|c|c|}
\hline
\multicolumn{4}{|c||}{} & \multicolumn{6}{c||}{Fields/Parameters}
& \multicolumn{5}{c|}{Antifields/Antiparameters} \\
\hline Quantity & $g$ & $\lambda$ & $x$ &
$A,\mathcal{B},\partial_{x}$ & $C,\chi$ & $\theta $ & $\xi$ & $B$ &
$\overline{C}$ & $A^{\ast },\mathcal{B}^{\ast} $ & $C^{\ast}$ &
$\theta^{\ast}$ & $\xi^{\ast}$ & $\overline{C}^{\ast}$
\\ \hline\hline $\varepsilon$   & 0 & 1 & 0 & 0 & 1 & 1 & 0 & 0 & 1
& 1 & 0 & 0 & 1 & 0  \\ \hline \textrm{gh}     & 0 &-1 & 0 & 0 &
1 & 1 & 0 & 0 &-1 &-1 &-2 &-2 &-1 & 0  \\ \hline
dim             & 0 &-1 &-1 & 1 & 1 & 2 & 0 & 2 & 1 & 2 & 2 & 1 & 3 & 2  \\ 
\hline
\end{tabular}
\end{center}
\caption{Grassmann parity $"\varepsilon "$,  ghost number $"\mathrm{
gh}"$, mass dimension $"\mathrm{dim}"$. }
\label{qnumbers}
\end{table}

\noindent The quantum numbers of any quantities encountered  in the
text are easily established using the quantum numbers in Table 1.
The  action that we construct is denoted by $S^{(1)}=S^{(1)}(\Phi
,\Phi^{\ast })$. We assume that the action has all the standard
properties,
\begin{equation}
\varepsilon (S^{(1)})=\mathrm{gh}(S^{(1)})=\dim (S^{(1)})=0,
\label{1.8a}
\end{equation}
and $S^{(1)}$ satisfies the master equation,
\begin{equation}
 S^{(1)}\!\!\int dx\; [\overleftarrow{\partial }_{\Phi }\partial _{\Phi ^{\ast }}]S^{(1)}=0,
 \qquad
\left. S^{(1)}\right| _{C,\mathcal{B},\theta ,\Phi ^{\ast }=0}=\mathcal{S}
_{YM}(A).  \label{1.8b}
\end{equation}
We also assume  that $S^{(1)}$ has a background gauge symmetry
\begin{equation}
S^{(1)}\overleftarrow{\tilde{h}^{\alpha }}\omega^{\alpha }=0,  \label{1.9}
\end{equation}
where
\begin{eqnarray}
\overleftarrow{\tilde{h}^{\alpha }}\omega^{\alpha }&=&\int dx\left\{ \left[
\frac{\overleftarrow{\delta }}{\delta \mathcal{B}_{\mu }^{\beta }}D_{\mu
}^{\beta \alpha }(\mathcal{B})+g\varepsilon ^{\beta \gamma \alpha }\left(
\frac{\overleftarrow{\delta }}{\delta A_{\mu }^{\beta }}A_{\mu }^{\gamma }+
\frac{\overleftarrow{\delta }}{\delta C^{\beta }}C^{\gamma }+\frac{
\overleftarrow{\delta }}{\delta \theta _{\mu }^{\beta }}\theta _{\mu
}^{\gamma }\right) \right. \right.   \notag \\
&&\left. +\left. g\varepsilon ^{\beta \gamma \alpha }\left( \frac{
\overleftarrow{\delta }}{\delta A_{\mu }^{\ast \beta }}A_{\mu }^{\ast \gamma
}+\frac{\overleftarrow{\delta }}{\delta C^{\ast \beta }}C^{\ast \gamma }+
\frac{\overleftarrow{\delta }}{\delta \mathcal{B}_{\mu }^{\ast \beta }}
\mathcal{B}_{\mu }^{\ast \gamma }+\frac{\overleftarrow{\delta }}{\delta
\theta_{\mu }^{\ast \beta }}\theta _{\mu }^{\ast \gamma }\right) \right]
\omega^{\alpha }\right\} .  \label{1.10}
\end{eqnarray}
Equation (\ref{1.9}) means that $S^{(1)}$ is required to be
invariant under gauge transformations of the  background field,
which is an important property of the background field method.

\noindent We represent the action in the form
\begin{equation}
S^{(1)}=S_{1}^{(1)}(A,\mathcal{B})+S_{\Phi ^{\ast }}^{(1)}(\Phi,\Phi^{\ast
}),\qquad \left. S_{\Phi ^{\ast }}^{(1)}\right\vert _{\Phi ^{\ast }=0}=0.
\end{equation}
It is easy to see that $S_{\Phi ^{\ast }}^{(1)}$ is linear in $\Phi^{\ast }$, i.e.,
\begin{equation}
S_{\Phi ^{\ast }}^{(1)}=S_{A^{\ast }}^{(1)}+S_{C^{\ast }}^{(1)}+S_{^{
\mathcal{B}^{\ast }}}^{(1)}+S_{\theta ^{\ast }}^{(1)},\label{1.14}
\end{equation}
where
\begin{eqnarray}
S_{A^{\ast }}^{(1)}
=\int dx\left[ A^{\ast }D(V)C\right] +\int dx\left[ A^{\ast }\theta
\right] .\label{1.15}
\end{eqnarray}
In deriving (\ref{1.15}) we use equality (\ref{1.9}). In what
follows, we  systematically use this method. For the remaining part
of   actions (\ref{1.14}), we thus obtain
\begin{equation}
S_{C^{\ast }}^{(1)}=Z_{2}\int dx\left[ C^{\ast \alpha }\frac{g}{2}
\varepsilon ^{\alpha \beta \gamma }C^{\beta }C^{\gamma }\right] ,
\label{1.16}
\end{equation}
\begin{equation}
S_{\mathcal{B}^{\ast }}^{(1)}=Z_{3}\int dx\left[ \mathcal{B}^{\ast }D(V)C
\right] -(Z_{4}-Z_{3})\int dx\left[ \mathcal{B}^{\ast }\theta \right] ,
\end{equation}
\begin{eqnarray}
&&S_{\theta ^{\ast }}^{(1)}=\sum_{k=1}^{3}S_{\theta ^{\ast
}k}^{(1)},\qquad S_{\theta ^{\ast }1}^{(1)}=Z_{5}\int dx\left[ \theta ^{\ast
}g\varepsilon \theta C\right] ,   \\
&&S_{\theta ^{\ast }2}^{(1)}=Z_{6}\int dx\left[ \theta ^{\ast }\varepsilon
C(D(V)C)\right] ,\qquad S_{\theta ^{\ast }3}^{(1)}=Z_{7}\int dx\left[ \theta ^{\ast
}g\varepsilon A(\varepsilon CC)\right] .  \label{1.18}
\end{eqnarray}
Let $A^{\ast }=\tilde{A}^{\ast }-Z_{3}\mathcal{B}^{\ast }$. Then
\begin{eqnarray}
&&S_{A^{\ast }}^{(1)}+S_{\mathcal{B}^{\ast }}^{(1)}=S_{\tilde{A}^{\ast
}}^{(1)}+\tilde{S}_{\mathcal{B}^{\ast }}^{(1)},   \\
&&S_{\tilde{A}^{\ast }}^{(1)}=S_{\tilde{A}^{\ast }1}^{(1)}+S_{\tilde{A}
^{\ast }2}^{(1)},\ S_{\tilde{A}^{\ast }1}^{(1)}=\int dx\left[ \tilde{A}
^{\ast }D(V)C\right] ,\qquad S_{\tilde{A}^{\ast }2}^{(1)}=\int dx\left[ \tilde{A}
^{\ast }\theta \right] ,  \label{1.19b} \\
&&\tilde{S}_{\mathcal{B}^{\ast }}^{(1)}=-Z_{4}\int dx\left[ \mathcal{B}
^{\ast }\theta \right] .  \label{1.19c}
\end{eqnarray}
Hence, $S_{\Phi ^{\ast }}^{(1)}$ has the form
\begin{eqnarray}
&&S_{\Phi ^{\ast }}^{(1)}=S_{\tilde{A}^{\ast }}^{(1)}+S_{C^{\ast }}^{(1)}+
\tilde{S}_{^{\mathcal{B}^{\ast }}}^{(1)}+S_{\theta ^{\ast }}^{(1)},
\label{1.20a} \\
&&S_{\tilde{A}^{\ast }}^{(1)}={\rm rhs}(\ref{1.19b}),\qquad
S_{C^{\ast }}^{(1)}={\rm rhs}(\ref{1.16}),\qquad
\tilde{S}_{\mathcal{B}^{\ast }}^{(1)}={\rm rhs}(\ref{1.19c}),\qquad
S_{\theta ^{\ast }}^{(1)}={\rm rhs}(\ref{1.18}).
\end{eqnarray}

\subsection{Master-equation}

\noindent We hereafter omit the tildes in (\ref{1.20a}). We consider
  consequences of  master equation (\ref{1.8b}). In what
follows,  we write different  anticanonical sectors of the left-hand
side of eq. (\ref{1.8b}) separately. The  action of each derivative
is labeled with an identification number $n$ for  convenience. The
identification number $n$ is written  as a subscript to a vertical
line in Eqs. (\ref{vert1})-(\ref{vert4}) below. We have
\begin{equation}
\left[ \left. S_{1}^{(1)}\overleftarrow{\partial }_{A^{\alpha }}\right\vert
_{1}+\left. S_{A^{\ast }1}^{(1)}\overleftarrow{\partial }_{A^{\alpha
}}\right\vert _{2}+\left. S_{\theta^{\ast}2}^{(1)}\overleftarrow{\partial }
_{A^{\alpha }}\right\vert _{16}+\left. S_{\theta^{\ast}3}^{(1)}\overleftarrow{
\partial }_{A^{\alpha }}\right\vert _{17}\right] \left[ \left. \partial
_{A^{\ast \alpha }}S_{A^{\ast }1}^{(1)}\right\vert _{3}+\left. \partial
_{A^{\ast \alpha }}S_{A^{\ast }2}^{(1)}\right\vert _{4}\right] ; \label{vert1}
\end{equation}
for $\overleftarrow{\partial }_{A}\partial _{A^{\ast }}$,
\begin{equation}
\left[ \left. S_{A^{\ast }1}^{(1)}\overleftarrow{\partial }_{C^{\alpha
}}\right\vert_{5}+\left. S_{C^{\ast }}^{(1)}\overleftarrow{\partial }
_{C^{\alpha }}\right\vert _{6}+\left. S_{\theta^{\ast}1}^{(1)} \overleftarrow{
\partial }_{C^{\alpha }}\right\vert_{7}+\left.S_{\theta^{\ast}2}^{(1)}
\overleftarrow{\partial }_{C^{\alpha
}}\right\vert_{18}+\left.S_{\theta^{\ast}3}^{(1)} \overleftarrow{\partial }
_{C^{\alpha }}\right\vert _{19}\right]\left[ \left. \partial _{C^{\ast
\alpha }}S_{C^{\ast }}^{(1) }\right\vert_{8}\right]; \label{vert2}
\end{equation}
for $\overleftarrow{\partial }_{C}\partial _{C^{\ast }}$,
\begin{equation}
\left[ \left. S_{1}^{(1)}\overleftarrow{\partial }_{\mathcal{B}^{\alpha
}}\right\vert_{9}+\left. S_{A^{\ast }1}^{(1)}\overleftarrow{\partial }_{
\mathcal{B}^{\alpha }}\right\vert _{10}+\left.S_{\theta^{\ast}2}^{(1)}
\overleftarrow{\partial }_{\mathcal{B}^{\alpha }}\right\vert _{20}\right]
\left[ \left. \partial _{\mathcal{B}^{\ast \alpha }}S_{\mathcal{B}^{\ast
}}^{(1)}=-Z_{4}\theta^{\alpha }\right\vert_{11}\right]; \label{vert3}
\end{equation}
for $\overleftarrow{\partial
}_{\mathcal{B}}\partial_{\mathcal{B}^{\ast }}$, and

 \noindent $\overleftarrow{\partial }_{\theta }\partial
_{\theta ^{\ast }}$ :
\begin{equation}
\left[ \left. S_{A^{\ast }2}^{(1)}\overleftarrow{\partial}_{\theta^{\alpha
}}\right\vert _{12}+\left. S_{\mathcal{B}^{\ast}}^{(1)}\overleftarrow{\partial }
_{\theta ^{\alpha }}\right\vert _{13}+\left. S_{\theta^{\ast}1}^{(1)}
\overleftarrow{\partial }_{\theta^{\alpha }}\right\vert_{14}\right] \left[
\left. \partial_{\theta^{{\ast}\alpha
}}S_{\theta^{\ast}1}^{(1)}\right\vert_{15}+\left. \partial_{\theta^{{\ast}\alpha
}}S_{\theta^{\ast}2}^{(1)}\right\vert_{21}+\left. \partial_{\theta^{{\ast}\alpha
}}S_{\theta^{\ast}3}^{(1)}\right\vert_{22}\right] .\label{vert4}
\end{equation}
for $\overleftarrow{\partial }_{\theta }\partial _{\theta ^{\ast}}$.
In what follows,  $n_1\bullet n_2$ denotes the antibracket
constructed from the action derivatives with the identification
numbers $n_1$ and $n_2$. The term proportional to
$\mathcal{B}^*\theta C$ has the form
\begin{eqnarray}
13\bullet 15&=&S_{\mathcal{B}^{\ast}}^{(1)}\overleftarrow{\partial}_
{\theta^{\alpha}}\partial_{\theta^{{\ast}\alpha}}S_{\theta^{\ast}1}^{(1)}=
-Z_{4}Z_{5}g\int dx\left[ \mathcal{B}^{\ast \alpha }\varepsilon^{\alpha
\beta \gamma }\theta ^{\beta }C^{\gamma }\right] =0
 \nonumber \\ & \Longrightarrow&\qquad
Z_{4}Z_{5}=0\qquad \Longrightarrow \qquad Z_{4}=0\ \mathrm{or}\ Z_{5}=0.
\end{eqnarray}
Let $Z_{4}=0$. We consider the term proportional to
$A^{k}\mathcal{B}^{l}\theta $ in Eq. (\ref{1.8b}) :
\begin{eqnarray}
0&=&1\bullet 4=\int dx\left[ S_{1}^{(1)}\overleftarrow{\partial }_{A^{\alpha
}}\ \theta ^{\alpha }\right]  \qquad\Longrightarrow\qquad
S_{1}^{(1)}\overleftarrow{\partial }_{A^{\alpha }}=0\nonumber \\
&\Longrightarrow&\qquad
Z_{4}\neq 0\ \Longrightarrow Z_{5}=0.
\end{eqnarray}
We now consider the term proportional to $\mathcal{B}^{\ast }$ in
Eq. (\ref{1.8b}). It follows from (\ref{1.8b}) that
\begin{eqnarray}
0&=&13\bullet (21+22)=\int dx\{S_{\mathcal{B}^{\ast }}^{(1)}\overleftarrow{
\partial }_{\theta ^{\alpha }}\partial _{\theta ^{\ast \alpha }}\left[
S_{\theta ^{\ast }2}^{(1)}+S_{\theta ^{\ast }3}^{(1)}\right] \}\nonumber \\
&=&\int dx\{Z_{4}\mathcal{B}^{\ast }[Z_{6}\theta ^{\ast }\varepsilon
C(D(V)C)+Z_{7}g\varepsilon A(\varepsilon CC)]\}
 \nonumber \\
& \Longrightarrow& \qquad
Z_{6}=Z_{7}=0\qquad \Longrightarrow \qquad S_{\theta ^{\ast }}^{(1)}=0.
\end{eqnarray}
The remaining  sectors of the left-hand side of Eq. (\ref{1.8b})
reduce to
\begin{equation}
\left[ \left. S_{1}^{(1)}\overleftarrow{\partial }_{A^{\alpha
}}\right| _{1}+\left. S_{A^{\ast }1}^{(1)}\overleftarrow{\partial
}_{A^{\alpha }}\right| _{2}\right] \left[ \left. \partial _{A^{\ast
\alpha }}S_{A^{\ast }1}^{(1)}\right| _{3}+\left. \partial _{A^{\ast
\alpha }}S_{A^{\ast }2}^{(1)}\right| _{4}\right]
\end{equation}
for $\overleftarrow{\partial }_{A}\partial _{A^{\ast }}$,
\begin{equation}
\left[ \left. S_{A^{\ast}1}^{(1)}\overleftarrow{\partial }_{C^{\alpha
}}\right\vert_{5}+\left. S_{C^{\ast }}^{(1)}\overleftarrow{\partial }
_{C^{\alpha }}\right\vert _{6}\right]\left[ \left. \partial _{C^{\ast \alpha
}}S_{C^{\ast }}^{(1) }\right\vert_{8}\right];
\end{equation}
for $\overleftarrow{\partial }_{C}\partial _{C^{\ast }}$, and
\begin{equation}
\left[ \left. S_{1}^{(1)}\overleftarrow{\partial }_{\mathcal{B}^{\alpha
}}\right\vert_{9}+\left. S_{A^{\ast }1}^{(1)}\overleftarrow{\partial }_{
\mathcal{B}^{\alpha }}\right\vert _{10}\right] \left[ \left. \partial _{
\mathcal{B}^{\ast \alpha }}S_{\mathcal{B}^{\ast
}}^{(1)}=-Z_{4}\theta^{\alpha }\right\vert_{11}\right].
\end{equation}
for $\overleftarrow{\partial
}_{\mathcal{B}}\partial_{\mathcal{B}^{\ast }}$ We obtain the term
proportional to $A^{\ast }C\theta $
\begin{equation}
2\bullet 4+10\bullet 11=0\qquad  \Longrightarrow\qquad  Z_{4}=1;
\end{equation}
and the term proportional to $A^{\ast }CD(V)C$ :
\begin{equation}
2\bullet 3+5\bullet 8=0=(Z_{2}-1)\frac{g}{2}\int dx\left[ A^{\ast \beta
}D^{\beta \alpha }(V)\varepsilon^{\alpha \delta \sigma }C^{\delta }C^{\sigma
}\right] \qquad \Longrightarrow\qquad  Z_{2}=1,
\end{equation}
where we use the identity
\begin{equation}
\varepsilon^{\alpha \sigma \gamma }\left[ D^{\sigma \beta
}(V)C^{\beta } \right] C^{\gamma }=\frac{1}{2}D^{\alpha \sigma
}(V)\left( \varepsilon^{\sigma \beta \gamma }C^{\beta }C^{\gamma
}\right).
\end{equation}
The terms proportional to $S_{1}^{(1)}\overleftarrow{\partial
}_{A^{\alpha }}\theta ^{\alpha }$ and
$S_{1}^{(1)}\overleftarrow{\partial }_{\mathcal{B}^{\alpha }}\theta
^{\alpha }$ are given by
\begin{equation}
1\bullet 4+9\bullet 11=0=S_{1}^{(1)}\int dx\left[ \left( \overleftarrow{
\partial }_{A^{\alpha }}-\overleftarrow{\partial }_{\mathcal{B}^{\alpha
}}\right) \theta ^{\alpha }\right]
\qquad  \Longrightarrow\qquad  S_{1}^{(1)}(A,\mathcal{B})=S_{1}^{(1)}(V);
\end{equation}
and the term proportional to $C^{\ast }CCC$ is given by
\begin{equation}
6\bullet 8=\frac{g^{2}}{2}\int dx\left( C^{\ast \beta }\varepsilon
^{\beta \gamma \alpha }C^{\gamma }\varepsilon ^{\alpha \sigma \delta
}C^{\sigma }C^{\delta }\right) \equiv 0.
\end{equation}
Finally, the terms depending only on the fields $V$ and $C$a are
represented by the formulas
\begin{equation}
0=1\bullet 3=S_{1}^{(1)}(V)\int dx\left[ \overleftarrow{\partial }_{V}D(V)C
\right]\qquad   \Longrightarrow\qquad  S_{1}^{(1)}(V)=S_{YM}(V),
\end{equation}
which agrees with the boundary condition of the master equation
(\ref{1.8b}).

\noindent We thus obtain the action as a solution to the master
equation with the given boundary condition
\begin{equation}
S^{(1)}=S_{YM}(V) +\int dx\left[ A^{\ast }D(V)C\right]
+\int dx\left[ C^{\ast \alpha }\frac{g}{2}\varepsilon ^{\alpha \beta
\gamma }C^{\beta }C^{\gamma }\right] +
\int dx\left[A^{\ast} \theta\right]-\int dx\left[\mathcal{B}^{\ast
} \theta\right] \label{minaction}.
\end{equation}
The first three  terms in  (\ref{minaction}) constitute the standard
minimal action in the BV formalism. The last two terms are expected
because the combination $A^{\ast}-\mathcal{B}^{\ast}$
antibracket-commutes with $V\equiv A+\mathcal{B}$.

\subsection{Anticanonical transformation}

\noindent We shift the action as
\begin{equation}
\ S^{(1)}\rightarrow \ S^{(2)}=S^{(1)}+\int dx\;\overline{C}^{\ast }B+\xi ^{\ast
}\chi ,
\end{equation}
and then  impose the gauge  via the anticanonical transformation
\begin{eqnarray}
&&\bar{\Phi}^{\ast }=Y(\bar{\Phi},\bar{\Phi}^{\ast \prime })
\overleftarrow{\partial }_{\bar{\Phi}},\qquad \bar{\Phi}^{\prime
}=\partial_{\bar{\Phi}^{\ast \prime }}Y(\bar{\Phi},\bar{\Phi}^{\ast
\prime }),  \label{1.39a} \\
&&Y(\bar{\Phi},\bar{\Phi}^{\ast \prime })=\int dx \;\bar{\Phi}^{\ast
\prime }\bar{\Phi}
+\Lambda (\bar{\Phi})\ \Longrightarrow   \label{1.39b} \\
&&\bar{\Phi}^{\prime }=\bar{\Phi},\qquad \bar{\Phi}^{\ast
}=\bar{\Phi}^{\ast \prime }+\Lambda (\bar{\Phi})
\overleftarrow{\partial }_{\bar{\Phi}},\   \label{1.39c} \\
&&\bar{\Phi}=\{A,C,\mathcal{B},\theta ,\overline{C},B,\xi ,\chi \},\qquad
\bar{\Phi}^{\ast }=\{A^{\ast },C^{\ast },\mathcal{B}^{\ast },\theta ^{\ast },
\overline{C}^{\ast },B^{\ast },\xi ^{\ast },\chi ^{\ast }\}.  \label{1.39d}
\end{eqnarray}
We choose $\Lambda (\bar{\Phi})$ in the form
\begin{eqnarray}
&&\Lambda (\bar{\Phi})=\int dx\;\overline{C}\left( D(\mathcal{B})A+\frac{\xi }{2}
B\right) \Longrightarrow  \\
&&A^{\ast }=A^{\ast \prime }-D(\mathcal{B})\overline{C},\qquad C^{\ast }=C^{\ast
\prime },\qquad \overline{C}^{\ast }=\overline{C}^{\ast \prime }+D(\mathcal{B})A+
\frac{\xi }{2}B, \\
&&\mathcal{B}^{\ast } =\mathcal{B}^{\ast \prime } +g\left(
\varepsilon A \overline{C}\right) ,\qquad \xi ^{\ast} =\xi
^{\ast\prime }+\frac{1}{2}\int dx\;\overline{ C}B.
\end{eqnarray}
Omitting primes, we obtain
\begin{eqnarray}
S_{\mathrm{ext}}&=&S_{YM}(V)+\int dx\left[ A^{\ast }D(V)C\right] +\int dx
\left[ A^{\ast }\theta \right] +\int dx\left[ \theta D(V)\overline{C}\right]
\nonumber \\
&&+\int dx\left[ \overline{C}D(\mathcal{B})D(V)C + BD(\mathcal{B})A + (\xi /2)BB
\right] -\frac{1}{2}\chi \int dx\left( \overline{C}B\right) \nonumber \\
&&+\int dx\left[ C^{\ast \alpha }\frac{g}{2}\varepsilon^{\alpha\beta\gamma}
C^{\beta }C^{\gamma }\right]
+\int dx\left[ \overline{C}^{\ast\alpha }B^{\alpha }\right]
-\int dx[\mathcal{B}^{\ast }\theta ]+\xi ^{\ast}\chi .
\end{eqnarray}
With the exception of the last three terms, this action is analogous
to the actions presented  by
 Kluberg-Stern and Zuber
\cite{K-SZ} and by Piguet and Sibold \cite{PS}.

\section{Action $S_{\mathrm{ext}}$ as solution of a set of equations}
\label{secssol}

\noindent In what follows,  we  often omit the integral symbol  to
have a more compact formula. The action $S_{\mathrm{ext}}$ satisfies
the follow set of equations:

1. Equations linear in functional and partial derivatives with
respect to fields and antifields and in partial derivatives with
respect to constant parameters,
\begin{eqnarray}
&&\partial_{\theta^{\ast}}S_{\mathrm{ext}}=\partial_{B^{\ast}}S_{\mathrm{ext}
}=\partial_{\chi^{\ast}} S_{\mathrm{ext}}=0,  \label{2.1a} \\
&&\partial_{\mathcal{B}^{\ast}}S_{\mathrm{ext}}=-\;\theta,\qquad \partial_{\xi ^{\ast
}} S_{\mathrm{ext}}=\chi,\qquad \partial_{\overline{C}^{\ast}}S_{\mathrm{ext}}=B,
\label{2.1b}
\end{eqnarray}
\begin{equation}
\partial _{B}S_{\mathrm{ext}}=D(\mathcal{B})A+\xi B-\frac{1}{2}\chi
\overline{C}+\overline{C}^{\ast },  \label{2.2}
\end{equation}
\begin{equation}
\left[ D^{\alpha \beta }(\mathcal{B})\partial _{A^{\ast \beta
}}-\partial _{ \overline{C}^{\alpha }}\right]
S_{\mathrm{ext}}=-g\varepsilon ^{\alpha \beta \gamma }A^{\beta
}\theta ^{\gamma }-\frac{1}{2}\chi B^{\alpha },  \label{2.3}.
\end{equation}

2. The master equation
\begin{equation}
S_{\mathrm{ext}}\overleftarrow{\partial
_{\bar{\Phi}}}\partial_{\bar{\Phi} ^{\ast }}S_{\mathrm{ext}}=0
\quad\Longrightarrow\quad S_{\mathrm{ext}}\overleftarrow{\partial
_{A}}\partial _{A^{\ast }}S_{
\mathrm{ext}}+S_{\mathrm{ext}}\overleftarrow{\partial _{C}}\partial
_{C^{\ast }}S_{\mathrm{ext}}+\left[ \chi \partial _{\xi }-\theta
\partial _{ \mathcal{B}}-B\partial _{\overline{C}}\right]
S_{\mathrm{ext}}=0. \label{2.4b}
\end{equation}

3. The background  gauge invariance
\begin{equation}
S_{\mathrm{ext}}\overleftarrow{H^{\alpha }}\omega^{\alpha }=0, \label{2.5}
\end{equation}
where
\begin{eqnarray}
\overleftarrow{H^{\alpha }}\omega^{\alpha }\!\!&=\!\!&\int \!\!dx\left\{ \left[ \frac{
\overleftarrow{\delta }}{\delta \mathcal{B}_{\mu }^{\beta }}D_{\mu }^{\beta
\alpha }(\mathcal{B})+g\varepsilon ^{\beta \gamma \alpha }\!\left( \frac{
\overleftarrow{\delta }}{\delta A_{\mu }^{\beta }}A_{\mu }^{\gamma }+\frac{
\overleftarrow{\delta }}{\delta C^{\beta }}C^{\gamma }+\frac{\overleftarrow{
\delta }}{\delta \theta _{\mu }^{\beta }}\theta _{\mu }^{\gamma }+\frac{
\overleftarrow{\delta }}{\delta \overline{C}^{\beta }}\overline{C}^{\gamma
}+\frac{
\overleftarrow{\delta }}{\delta B^{\beta }}B^{\gamma }\right) \right. \right.   \notag \\
&&\left. +\left. g\varepsilon ^{\beta \gamma \alpha }\left( \frac{
\overleftarrow{\delta }}{\delta A_{\mu }^{\ast \beta }}A_{\mu }^{\ast \gamma
}+\frac{\overleftarrow{\delta }}{\delta C^{\ast \beta }}C^{\ast \gamma }+
\frac{\overleftarrow{\delta }}{\delta \mathcal{B}_{\mu }^{\ast \beta }}
\mathcal{B}_{\mu }^{\ast \gamma }+\frac{\overleftarrow{\delta }}{\delta
\theta _{\mu }^{\ast \beta }}\theta _{\mu }^{\ast \gamma }+\frac{
\overleftarrow{\delta }}{\delta \overline{C}^{\ast \beta }}\overline{C}
^{\ast \gamma }\right) \right] \omega^{\alpha }\right\} .\qquad  \label{2.6}
\end{eqnarray}

\section{General solution of equation system (\protect\ref{2.1a}) - (\protect
\ref{2.6})}
\label{secpsol}

\noindent We  find a functional $P=P(\bar{\Phi},\bar{\Phi}^{\ast })$
that has all quantum numbers of the functional $S_{\mathrm{ext}}$
and satisfies  system of equation (\ref{2.1a}) - (\ref{2.6}) (with
substitution $S_{\mathrm{ext }}\rightarrow P$)
\begin{eqnarray}
&&\partial_{\theta^{\ast}}P=\partial_{B^{\ast}}P=\partial_{\chi^{\ast}}P=0,
\label{3.1a} \\
&&\partial_{\mathcal{B}^{\ast}}P=-\theta, \qquad \partial_{\overline{C}^{\ast}}P=B, \qquad
\partial_{\xi ^{\ast }}P=\chi ,  \label{3.1b}
\end{eqnarray}
\begin{equation}
\partial _{B}P=D(\mathcal{B})A+\xi B-\frac{1}{2}\chi \overline{C}+\overline{C
}^{\ast },  \label{3.2}
\end{equation}
\begin{equation}
\left[ D^{\alpha \beta }(\mathcal{B})\partial _{A^{\ast \beta }}-\partial _{
\overline{C}^{\alpha }}\right] P=-g\varepsilon ^{\alpha \beta \gamma
}A^{\beta }\theta ^{\gamma }-\frac{1}{2}\chi B^{\alpha },  \label{3.3}
\end{equation}
the master equation
\begin{equation}
P\overleftarrow{\partial _{\bar{\Phi}}}\partial _{\bar{\Phi}^{\ast
}}P=0\qquad\Longrightarrow\qquad
P\overleftarrow{\partial _{A}}\partial _{A^{\ast }}P+P\overleftarrow{
\partial _{C}}\partial _{C^{\ast }}P+\left[ \chi \partial _{\xi }-\theta
\partial _{\mathcal{B}}-B\partial _{\overline{C}}\right] P=0,  \label{3.4}
\end{equation}
and the background gauge invariance
\begin{equation}
P\overleftarrow{H^{\alpha }}\omega^{\alpha }=0,  \label{3.5a}
\end{equation}
where the operator $\overleftarrow{H^{\alpha }}\omega^{\alpha }$
 is given by Eq. (\ref{2.6}).

\noindent It follows from Eqs. (\ref{3.1a}) that the functional
$P(\bar{\Phi},\bar{\Phi} ^{\ast })$ is independent of the antifields
$\theta ^{\ast }$ and  $B^{\ast } $ and fermionic parameter
$\chi^{\ast }$. Because $\theta^{\ast }$, $B^{\ast }$ and $ \chi
^{\ast }$ do not appear in what follows, we understand  the set
$\bar{\Phi }^{\ast }$ as the set $\bar{\Phi}^{\ast }=\{A^{\ast
},C^{\ast },\mathcal{B} ^{\ast },\overline{C}^{\ast },\xi ^{\ast
}\}$. The operator $\overleftarrow{ H^{\alpha }}\omega^{\alpha }$
reduces to
\begin{eqnarray}
\overleftarrow{H^{\alpha }}\omega^{\alpha }\!\!&=\!\!&\int dx\left\{ \left[ \frac{
\overleftarrow{\delta }}{\delta _{\mu }^{\beta }}D_{\mu }^{\beta \alpha }(
\mathcal{B})+g\varepsilon ^{\beta \gamma \alpha }\left( \frac{\overleftarrow{
\delta }}{\delta A_{\mu }^{\beta }}A_{\mu }^{\gamma }+\frac{\overleftarrow{
\delta }}{\delta C^{\beta }}C^{\gamma }+\frac{\overleftarrow{\delta }}{
\delta \theta _{\mu }^{\beta }}\theta _{\mu }^{\gamma }+\frac{\overleftarrow{
\delta }}{\delta \overline{C}^{\beta }}\overline{C}^{\gamma }+\frac{
\overleftarrow{\delta }}{\delta B^{\beta }}B^{\gamma }\right) \right.
\right.   \notag \\
&&\left. +\left. g\varepsilon ^{\beta \gamma \alpha }\left( \frac{
\overleftarrow{\delta }}{\delta A_{\mu }^{\ast \beta }}A_{\mu }^{\ast \gamma
}+\frac{\overleftarrow{\delta }}{\delta C^{\ast \beta }}C^{\ast \gamma }+
\frac{\overleftarrow{\delta }}{\delta \mathcal{B}_{\mu }^{\ast \beta }}
\mathcal{B}_{\mu }^{\ast \gamma }+\frac{\overleftarrow{\delta }}{\delta
\overline{C}^{\ast \beta }}\overline{C}^{\ast \gamma }\right) \right]
\omega^{\alpha }\right\} .  \label{3.5b}
\end{eqnarray}
We write $P$ in the form
\begin{eqnarray}
P&=&P_{00}+\check{P},\qquad \check{P}=P^{(1)}+\chi P^{(2)}, \\
P_{00}&=&\int dx[\overline{C}^{\ast}B
-\mathcal{B}^{\ast}\theta +BD(\mathcal{B})A +(\xi /2)BB\nonumber \\
&&+g\overline{C}\varepsilon A\theta ]
-\frac{1}{2}\chi \int dx[\overline{C}B] +\xi ^{\ast }\chi .
\end{eqnarray}
We note that $P_{00}$ satisfies the equation of the type
(\ref{3.5a}),
\begin{equation}
P_{00}\overleftarrow{H^{\alpha }}\omega^{\alpha }=0.  \label{3.6}
\end{equation}
It follows from Eqs. (\ref{3.1b}) and (\ref{3.2})
\begin{equation}
\partial _{\mathcal{B}^{\ast }}P^{(k)}=\partial _{\overline{C}^{\ast
}}P^{(k)}=\partial _{\xi ^{\ast }}P^{(k)}=\partial _{B}P^{(k)}=0,\qquad k=1,2,
\label{3.7}
\end{equation}
and hence
\begin{eqnarray}
P^{(k)} &=&P^{(k)}\left( \tilde{\Phi},\Phi ^{(1)\ast }\right)
,\qquad \tilde{\Phi} =\{\Phi ^{(1)},\Phi ^{(2)}\}, \label{3.8a}
\\
 \Phi ^{(1)}&=&\left\{ A,C\right\}
,\qquad \Phi^{(2)}=\left\{ \mathcal{B},\theta ,\overline{C},\xi ,\chi \right\}
,\qquad \Phi^{(1)\ast }=\left\{ A^{\ast },C^{\ast }\right\} .  \label{3.8b}
\end{eqnarray}
It follows from Eqs. (\ref{3.3})
\begin{equation}
\left[ D^{\alpha \beta }(\mathcal{B})\partial _{A^{\ast \beta }}-\partial _{
\overline{C}^{\alpha }}\right] P^{(k)}=0,\qquad k=1,2.  \label{3.9}
\end{equation}
Hence, the functionals $P^{(k)}$ satisfy Eq. (\ref{3.9}), the
equations
\begin{equation}
P^{(k)}\overleftarrow{h^{\alpha }}\omega^{\alpha }=0,
\qquad k=1,2,  \label{3.10}
\end{equation}
\begin{eqnarray}
\overleftarrow{h^{\alpha }}\omega^{\alpha }&=&\int dx\left\{ \left[ \frac{
\overleftarrow{\delta }}{\delta \mathcal{B}_{\mu }^{\beta }}D_{\mu }^{\beta
\alpha }(\mathcal{B})+g\varepsilon ^{\beta \gamma \alpha }\left( \frac{
\overleftarrow{\delta }}{\delta A_{\mu }^{\beta }}A_{\mu }^{\gamma }+\frac{
\overleftarrow{\delta }}{\delta C^{\beta }}C^{\gamma }+\frac{\overleftarrow{
\delta }}{\delta \theta _{\mu }^{\beta }}\theta _{\mu }^{\gamma }+\frac{
\overleftarrow{\delta }}{\delta \overline{C}^{\beta }}\overline{C}^{\gamma
}\right) \right. \right.   \notag \\
&&\left. +\left. g\varepsilon ^{\beta \gamma \alpha }\left( \frac{
\overleftarrow{\delta }}{\delta A_{\mu }^{\ast \beta }}A_{\mu }^{\ast \gamma
}+\frac{\overleftarrow{\delta }}{\delta C^{\ast \beta }}C^{\ast \gamma
}\right) \right] \omega^{\alpha }\right\} ,  \label{3.11a}
\end{eqnarray}
and the equations  following from Eq. (\ref{3.4}).

\noindent We write  the functionals $P^{(k)}$ in the form
\begin{equation}
P^{(k)}(A^{\ast },\overline{C},\Psi )=\tilde{P}^{(k)}(\mathcal{A}^{\ast },
\overline{C},\Psi ),
\end{equation}
where $\Psi $ is the set of all variables in  the functionals
$P^{(k)}$ in addition to $A^{\ast },\overline{C}$,
\begin{equation}
\Psi =\{A,C,\mathcal{B},\theta ,\xi ,C^{\ast }\},
\end{equation}
and we introduce the expression for  $\mathcal{A}^{\ast }$
\begin{equation}
\mathcal{A}^{\ast }=A^{\ast }-D(\mathcal{B)}\overline{C}.  \label{3.13}
\end{equation}
It follows from Eq. (\ref{3.9}) that
\begin{equation}
\partial_{\overline{C}^{\alpha }}\left. \tilde{P}^{(k)}(\mathcal{A}
^{\ast },\overline{C},\Psi )\right\vert_{\mathcal{A}^{\ast },\Psi
}=0 \qquad\Longrightarrow\qquad
P^{(k)}(A^{\ast },\overline{C},\Psi )=\tilde{P}^{(k)}(\mathcal{A}^{\ast
},\Psi ),\qquad k=1,2,
\end{equation}
where $\mathcal{A}^{\ast }$ is given by Eq. (\ref{3.13}).

\subsection{A shift}

\noindent We change the  variables $A_{\mu }^{\ast \alpha }$,
$\mathcal{B} _{\mu }^{\alpha }$ and $\overline{C}^{\alpha }$,
\begin{equation}
A^{\ast }=A^{\ast ^{\prime }}+D(\mathcal{B)}\overline{C}^{\prime },
\qquad \mathcal{B}=\mathcal{B
}^{\prime },\qquad \overline{C}=\overline{C}^{\prime },
\end{equation}
\begin{eqnarray}
&\partial _{A^{\ast }}F=&\partial _{\mathcal{A}^{\ast }}\tilde{F},\ \partial
_{\overline{C}}F=\left[ \partial _{\overline{C}^{\prime }}+D(\mathcal{B}
^{\prime })\partial _{\mathcal{A}^{\ast }}\right] \tilde{F}, \\
&\partial _{\mathcal{B}}F=&\left[ \partial _{\mathcal{B}^{\prime
}}-g\varepsilon \overline{C}^{\prime }\partial _{\mathcal{A}^{\ast }}\right]
\tilde{F},
\end{eqnarray}
\begin{eqnarray}
&\partial _{\mathcal{A}^{\ast }}\tilde{F}=&\partial _{A^{\ast }}F,\ \partial
_{\overline{C}^{\prime }}\tilde{F}=\left[ \partial _{\overline{C}}-D(
\mathcal{B})\partial _{A^{\ast }}\right] F, \\
&\partial _{\mathcal{B}^{\prime }}\tilde{F}=&\left[ \partial _{\mathcal{B}
}+g\varepsilon \overline{C}\partial _{A^{\ast }}\right] F,
\end{eqnarray}
\begin{equation}
P^{(k)}\int dx \left[ -\overleftarrow{\partial _{\mathcal{B}}}\overleftarrow{D(
\mathcal{B)}}+\overleftarrow{\partial _{A^{\ast }}}A^{\ast }\varepsilon -
\overleftarrow{\partial _{\overline{C}}}\overline{C}\varepsilon \right]
\omega
=\tilde{P}^{(k)}\int dx \left[ -\overleftarrow{\partial _{\mathcal{B}^{\prime }}}
\overleftarrow{D(\mathcal{B}^{\prime }\mathcal{)}}+\overleftarrow{\partial _{
\mathcal{A}^{\ast }}}\mathcal{A}^{\ast }\varepsilon \right] \omega ,
\end{equation}
where $\mathcal{A}^{\ast }$ is regarded  as an independent variable
under the action on the functional $\tilde{F}$, $\mathcal{A}^{\ast
}\equiv A^{\ast ^{\prime }}$, and
\begin{equation}
F=F(A^{\ast },\mathcal{B},\overline{C},\ldots)=\tilde{F}=\tilde{F}(\mathcal{A}
^{\ast },\mathcal{B}^{\prime },\overline{C}^{\prime },\ldots),
\end{equation}
where ellipsis ``$\ldots$'' means all the remaining invariable arguments.

Equations  (\ref{3.10}) and (\ref{3.11a})  reduce to (omitting
primes)
\begin{equation}
\tilde{P}^{(k)}\overleftarrow{\tilde{h}^{\alpha }}\omega^{\alpha }=0,
\qquad k=1,2,  \label{3.15}
\end{equation}
\begin{eqnarray}
\overleftarrow{\tilde{h}^{\alpha }}\omega^{\alpha }&=&\int dx\left\{ \left[
\frac{\overleftarrow{\delta }}{\delta \mathcal{B}_{\mu }^{\beta }}D_{\mu
}^{\beta \alpha }(\mathcal{B})+g\varepsilon ^{\beta \gamma \alpha }\left(
\frac{\overleftarrow{\delta }}{\delta A_{\mu }^{\beta }}A_{\mu }^{\gamma }+
\frac{\overleftarrow{\delta }}{\delta C^{\beta }}C^{\gamma }+\frac{
\overleftarrow{\delta }}{\delta \theta _{\mu }^{\beta }}\theta _{\mu
}^{\gamma }\right) \right. \right.   \notag \\
&&\left. +\left. g\varepsilon ^{\beta \gamma \alpha }\left( \frac{
\overleftarrow{\delta }}{\delta \mathcal{A}_{\mu }^{\ast \beta }}\mathcal{A}
_{\mu }^{\ast \gamma }+\frac{\overleftarrow{\delta }}{\delta C^{\ast \beta }}
C^{\ast \gamma }\right) \right] \omega^{\alpha }\right\} .  \label{3.16a}
\end{eqnarray}
It follows from Eq. (\ref{3.4}) that the equations for
$\tilde{P}^{(k)}$ are (omitting primes),
\begin{eqnarray}
&&\tilde{P}^{(1)}\overleftarrow{\partial _{\mathcal{A}}}\partial _{\mathcal{A
}^{\ast }}\tilde{P}^{(1)}+\tilde{P}^{(1)}\overleftarrow{\partial _{\mathcal{C
}}}\partial _{C^{\ast }}\tilde{P}^{(1)}-\theta \partial _{\mathcal{B}}\tilde{
P}^{(1)}=0,  \label{3.17a} \\
&&\ \partial _{\xi }\tilde{P}^{(1)}=\tilde{P}^{(1)}\overleftarrow{\partial _{
\mathcal{A}}}\partial _{\mathcal{A}^{\ast }}\tilde{P}^{(2)}-\tilde{P}^{(2)}
\overleftarrow{\partial _{\mathcal{A}}}\partial _{\mathcal{A}^{\ast }}\tilde{
P}^{(1)}  \notag \\
&&+\tilde{P}^{(1)}\overleftarrow{\partial _{\mathcal{C}}}\partial _{C^{\ast
}}\tilde{P}^{(2)}-\tilde{P}^{(2)}\overleftarrow{\partial _{\mathcal{C}}}
\partial _{C^{\ast }}\tilde{P}^{(1)}-\theta \partial _{\mathcal{B}}\tilde{P}
^{(2)},\qquad \mathcal{A}\equiv A.  \label{3.17b}
\end{eqnarray}

\subsection{Solving of Eq. (\protect\ref{3.17a})}

\noindent Eq. (\ref{3.17a}) for $\tilde{P}^{(1)}$ was already solved
in \cite{BLT-YM1}(see   Eq. (3.51) in Sec. 3.1.) The result (with
 spinor arguments in Ref. \cite{BLT-YM1} set equal to zero) is
\begin{eqnarray}
&&\tilde{P}^{(1)}=\int dx
\left[-\frac{1}{4}Z_{14}G(U)G(U)-Z_{5}\theta \mathcal{A}^{\ast
}+Z_{6}\mathcal{A}^{\ast }D(U)C
+\frac{Z_{6}}{Z_{5}}\frac{g}{2}C^{\ast }\varepsilon CC\right] , \\
&& U=\mathcal{B}+\frac{1}{Z_{5}}\mathcal{A},\qquad Z_{5},Z_{6},Z_{14}\neq 0.
\end{eqnarray}

\subsection{Solving of Eq. (\protect\ref{3.17b})}

\noindent We first find an explicit form of the functional
$\tilde{P}^{(2)}$. The set of quantum numbers of $\tilde{P}^{(2)}$
and Eqs. (\ref{3.15}) give
\begin{equation}
\tilde{P}^{(2)}=\int dx\left[ Z_{1}\mathcal{A}^{\ast }\mathcal{A}
+Z_{2}C^{\ast }C\right] .
\end{equation}
Equation  (\ref{3.17b}) then  reduces to
\begin{eqnarray}
&&\ \partial _{\xi }\tilde{P}^{(1)}-\mathcal{L}\tilde{P}^{(1)}=0,
\label{3.2.2a} \\
&&\mathcal{L}=\int dx\left[ Z_{1}\left( \mathcal{A}\partial _{\mathcal{A}}-
\mathcal{A}^{\ast }\partial _{\mathcal{A}^{\ast }}\right) +Z_{2}\left(
C\partial _{C}-C^{\ast }\partial _{C^{\ast }}\right) \right] .
\end{eqnarray}
We write the functional $\partial_{\xi
}\tilde{P}^{(1)}-\mathcal{L}\tilde{P} ^{(1)}$ as a linear
combination of independent monomials $V_{k}(
\mathcal{A},C,\mathcal{A}^{\ast },C^{\ast },\theta )$ with
coefficients (differential operators) $M_{k}(\mathcal{B})$,
\begin{eqnarray}
&&\ \partial _{\xi }\tilde{P}^{(1)}-\mathcal{L}\tilde{P}^{(1)}=\sum_{k}M_{k}(
\mathcal{B})V_{k}(\mathcal{A},C,\mathcal{A}^{\ast },C^{\ast },\theta ), \\
&&M_{1}V_{1}=m_{1}\int dx[\mathcal{A}^{\ast }\theta ],\qquad m_{1}=\dot{Z}
_{5}-Z_{1}Z_{5}, \\
&&M_{2}V_{2}=m_{2}\int dx[\frac{g}{2}C^{\ast }\varepsilon CC],\qquad m_{2}=\left(
\frac{Z_{6}}{Z_{5}}\right) ^{\cdot }-Z_{2}\frac{Z_{6}}{Z_{5}}, \\
&&M_{3}V_{3}=m_{3}\int dx[-\frac{1}{4}G(\mathcal{B})G(\mathcal{B})].\qquad m_{3}=
\dot{Z}_{14}, \\
&&\left. M_{k}\right\vert _{m_{1}=m_{2}=m_{3}=0}=0,\qquad k\geq 4.
\end{eqnarray}
It follows from Eq. (\ref{3.2.2a}) that (a dot over a constant
denotes the derivatives with respect to the parameter $\xi$)
\begin{eqnarray}
m_{1} &=&0\ \Longrightarrow Z_{1}=\frac{\dot{Z}_{5}}{Z_{5}},\qquad m_{2}=0\
\Longrightarrow Z_{2}=\frac{\dot{Z}_{6}}{Z_{6}}-\frac{\dot{Z}_{5}}{Z_{5}}, \\
m_{3} &=&0\ \Longrightarrow \dot{Z}_{14}=0. \label{Z14}
\end{eqnarray}
If $Z_{\ell}$ are the coefficients in counterterms, then
$Z_{\ell}=Z_{\ell}(\eta)$ can be represented as a Taylor series,
\begin{equation}
Z_{\ell}=\sum_{n=0}^{\infty }\eta ^{n}z_{\ell,n},
\end{equation}
where $z_{\ell,n}$ are formed from $n$-loop diagrams. In the tree
approximation, we then obtain
\begin{eqnarray}
Z_{5,0} &=&1,\qquad \dot{Z}_{5,0}=0,\qquad Z_{6,0}=1,\qquad \dot{Z}_{6,0}=0,
\nonumber \\
Z_{1,0} &=&\dot{Z}_{5,0}=0,\qquad Z_{2,0}=\dot{Z}_{6,0}-\dot{Z}_{5,0}=0,
\end{eqnarray}
i.e., the vertices $\chi \int dx[\mathcal{A}^{\ast }\mathcal{A]}$
and $ \chi \int dx\left[ C^{\ast }C\right] $ are absent in the tree
approximation.

\section{Summary}

\noindent We have found that constructing a renormalized action
using the background field method in the framework of the BV
formalism leads to a violation  of (exact) multiplicativity. But, it
does not lead to any difficulties if we are interested in the
physical sector. In the considered model,  we obtained a theory with
the  renormalized action $S_{ext}\!=\!P$ and can make the theory
finite  using the standard scheme for introducing counterterms. If
we set $A^{*}\!=\!C^{*}\!=\!0$, then we obtain  a sector where the
renormalization is already multiplicative and completely contains
the  physical sector. In particular, the gauge independence of the
renormalization constant $Z_{14}$ is preserved (see Eq. (\ref{Z14}).
We can say that  theories with this property  are {\it
quasimultiplicative renormalizable}.

\section*{Acknowledgments}

\noindent The work of I.A. Batalin and I.V. Tyutin is supported in part by the RFBR
grant 17-02-00317. The work of P.M. Lavrov is
supported partially by the Ministry of Science and Higher Education of
the Russian Federation, grant 3.1386.2017 and by the RFBR grant 18-02-00153.

\end{document}